\documentclass[english,twocolumn,prl]{revtex4-2}
\usepackage[T1]{fontenc}
\usepackage[latin9]{inputenc}
\usepackage{amsmath}
\usepackage{amssymb}
\usepackage{graphicx}

\makeatletter
\usepackage{hyperref}
\hypersetup{colorlinks=true,citecolor=blue,linkcolor=blue, urlcolor=blue, breaklinks=true}

\makeatother

\usepackage{babel}
\begin{document}
\title{Negative superfluid density and spatial instabilities in driven superconductors}
\author{Andrey Grankin and Victor Galitski}
\affiliation{Joint Quantum Institute, Department of Physics, University of Maryland,
College Park, MD 20742, USA}
\author{Vadim Oganesyan}
\affiliation{Physics Program and Initiative for the Theoretical Sciences, The Graduate Center, CUNY, New York, New York 10016, USA; 
Department of Physics and Astronomy, College of Staten Island, CUNY, Staten Island, New York 10314, USA}
\begin{abstract}
We consider excitation of Higgs modes via the modulation of the BCS
coupling within the Migdal-Eliashberg-Keldysh theory of time-dependent
superconductivity. Despite the presence of phonons, which break integrability,
we observe Higgs amplitude oscillations reminiscent of the integrable
case. The dynamics of quasiparticles follows from the effective Bogolyubov-de
Gennes equations, which represent a Floquet problem for the Bogolyubov
quasiparticles. We find that when the Floquet-Bogolyubov bands overlap,
the homogeneous solution formally leads to a negative superfluid density,
which is no longer proportional to the amplitude of the order parameter.
This result indicates an instability, which we explore using spatially-resolved
BdG equations. Spontaneous appearance of spatial inhomogeneities in
the order parameter is observed and they first occur when the superfluid
density becomes unphysical. We conclude that the homogeneous solution
to time-dependent superconductivity is generally unstable and breaks
up into a complicated spatial landscape via an avalanche of topological
excitations. 
\end{abstract}
\maketitle

Magnetic field expulsion \citep{MO33} is one of the hallmarks of
superconductivity. It originates from the diamagnetic supercurrent
${\bf j}_{s}$ which is related to the gauge potential ${\bf A}$
via London's equation ${\bf j}_{s}=-\frac{n_{s}}{m}{\bf A}$, where
$m$ is the effective electron mass and $n_{s}$ is the superfluid
density. The latter also characterizes the stiffness of phase fluctuations
$\theta$ in superfluids with the classical free energy being given
by $\delta F\propto n_{s}\int d{\bf r}\left(\nabla\theta\right)^{2}$
\citep{AS10}. The thermodynamic stability of bulk samples requires
$n_{s}>0$ and the negative values of the superfluid density resulting
in a paramagnetic Meissner response, have been associated with the
instability of the system towards the formation of spatial inhomogeneities.
Well-known examples include the Fulde-Ferrell \citep{FF64} and Larkin-Ovchinnikov
\citep{L64} states, which occur in superconductors in the presence
of a magnetic field or spin imbalance in neutral superfluids. Negative
values of $n_{s}$ were also theoretically predicted for the odd-frequency
superconducting states, implying they are thermodynamically unstable
\citep{CMT93}. Many such scenarios are characterized by emergent
Bogolyubov Fermi surfaces, that possess an enhanced density of states
of quasiparticles. This results in large paramagnetic contribution
to the electromagnetic response, thereby rendering $n_{s}$ negative. 

In this Letter, we demonstrate that negative values of $n_{s}$ can
occur in conventional superconductors in the presence  amplitude (Higgs) oscillations of the order parameter, which an implies instability of the homogeneous solution.
Different proposals for generation of Higgs oscillations were extensively
studied in the past and include interaction quenches \citep{YTA06,BLS04,GG23}
and an ultrafast terahertz pumping \citep{KSM15,MTF14}. Oscillatory
behavior of the superfluid density in the presence of the Higgs excitations
was shown to lead to the parametric amplification/generation of photons
\citep{BJC21,DYA09}. In this Letter, we demonstrate that in addition
to the oscillatory behavior, the superfluid density can acquire a divergent
negative static contribution. We attribute this component to the emergent
Floquet-Bogolyubov bands, in an ideal case where the electron-hole
recombination is impossible \citep{CEK23}, characterized by a divergent
density of states (DOS).  Interaction with phonons broadens the DOS, reducing this singularity. Finally, we simulate a quasi-1-dimensional disordered
superconductor and observe proliferation of solitons, concurrently
with $n_{s}$ acquiring negative values in a uniform case. 

\begin{figure}
\includegraphics[scale=0.25]{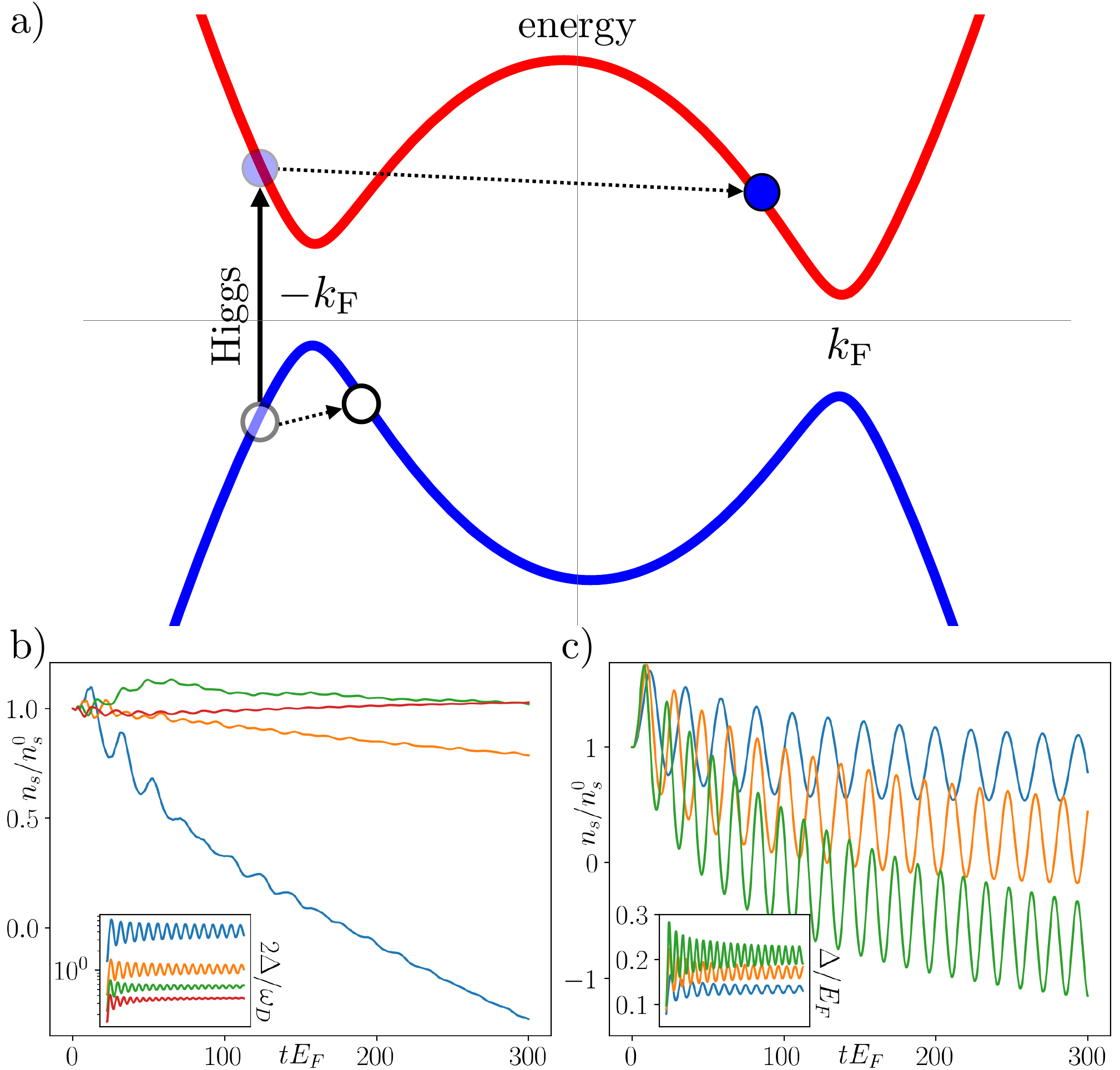}

\caption{Superfluid density in a quenched superconductor. (a) Schematic representation
of the quasiparticle dispersion in a superconductor in the presence
of external magnetic field. Arrows represent the scattering processes
leading to the generation of incoherent quasiparticle population.
(b) clean superconductor with electron-phonon interaction with $\lambda_{i}=0.25$
and $\lambda_{f}=0.45$ and different values of the Debye energy $\omega_{\text{D}}$:
$0.04E_{F}$ (blue), 0.16 (orange), 0.35 (green), 0.6 (red). (c) Disordered
case, quench parameters $\lambda_{i}=0.3$ and $\lambda_{f}$ being
equal to $0.4$ (blue), $0.45$ (orange), $0.5$ (green).}

\label{Fig1}
\end{figure}

We start by numerically evaluating the superfluid density within the
BCS-Holstein model undergoing a rapid change of the interaction
strength. The full Hamiltonian reads $H\left(t\right)=H_{\text{BCS}}\left(t\right)+H_{\text{el}-\text{ph}}+H_{\text{dis}}$:

\begin{align}
H_{\text{BCS}}=\sum_{k,\sigma}\xi_{{\bf k}}\psi_{{\bf k},\sigma}^{\dagger}\psi_{{\bf k},\sigma}-\lambda\left(t\right)\nu_{0}^{-1}\sum_{{\bf q}}\varrho_{{\bf q}}\varrho_{-{\bf q}}\label{eq:Hinst}\\
H_{\text{dis}}=\frac{1}{\sqrt{V}}\sum_{{\bf k,q},\sigma}U_{{\bf q}}\psi_{{\bf k}-{\bf q},\sigma}^{\dagger}\psi_{{\bf k},\sigma},\\
H_{\text{el-ph}}=\sum_{{\bf q}}g_{{\bf q}}\phi_{{\bf q}}\varrho_{-{\bf q}}+\sum_{{\bf q}}\omega_{{\bf q}}a_{{\bf q}}^{\dagger}a_{{\bf q}},
\end{align}
where $\varrho_{{\bf q}}=V^{-1/2}\sum_{{\bf k},\sigma}\psi_{{\bf k}+{\bf q},\sigma}^{\dagger}\psi_{{\bf k},\sigma}$,
$\phi_{{\bf q}}=a_{{\bf q}}+a_{-{\bf q}}^{\dagger}$, $\psi_{{\bf k},\sigma}$($\psi_{{\bf k},\sigma}^{\dagger}$)
and $a_{{\bf q}}(a_{{\bf q}}^{\dagger})$ respectively denote the
electron and phonon annihilation (creation). Their dispersions are
respectively defined as $\xi_{{\bf k}}=k^{2}/2m-E_{F}$ and $\omega_{q}$, where $m$ is the electronic mass and $E_{F}$ is the Fermi energy.
$g$ is the electron-phonon coupling strength and $V$ is the volume
of the system, $\nu_{0}$ is the electron density of states at the
Fermi level, $\lambda$ is the strength of short-range contact interaction representing an interaction with a high-energy phonon band.
$U_{{\bf q}}$ denotes Fourier transform of the local disorder potential.
Our model includes both a local BCS pairing interaction, characterized by the constant
$\lambda$, and phonons, which play the role of a bath leading to thermalization.
The interaction with the electromagnetic field is included via the minimal substitution ${\bf k\rightarrow k-A}$
in the electronic dispersion.

\paragraph{Quench dynamics}

In order to induce the amplitude oscillations in our model, we consider
a rapid change in the BCS pairing constant at $t=0$ from $\lambda\left(0^{-}\right)=\lambda_{i}$
to $\lambda\left(0^{+}\right)=\lambda_{f}$. We describe the time
evolution within the mean-field and disorder-averaged Gor'kov equations
 for the non-equilibrium Green's function, defined on the Kadanoff-Baym
contour \citep{Nessi}:
\begin{equation}
\left(i\partial_{t}-\hat{h}_{{\bf k}}\left(t\right)-\hat{\Sigma}_{{\bf k}}\star\right)\hat{G}_{{\bf k}}\left(t,t'\right)=\delta\left(t,t'\right)\hat{\mathbb{I}},\label{eq:G}
\end{equation}
where $\hat{G}_{{\bf k}}$ is the contour-ordered Green's function
defined as $\hat{G}_{{\bf k}}\left(t,t'\right)=-i\left\langle T_{{\cal C}}\Psi_{{\bf k}}\left(t\right)\otimes\Psi_{{\bf k}}^{\dagger}\left(t'\right)\right\rangle $,
with $\Psi_{{\bf k}}=\left(\begin{array}{c}
\psi_{{\bf k},\uparrow}\\
\psi_{-{\bf k},\downarrow}^{\dagger}
\end{array}\right)$ denoting Nambu spinors. $\hat{h}_{{\bf k}}\left(t\right)=\xi_{{\bf k}}\hat{\tau}_{3}+\Delta\left(t\right)\hat{\tau}_{1}$
is the effective Bogolyubov-de Gennes Hamiltonian (BdG) with $\Delta\left(t\right)$
being the instantaneous part of the anomalous self-energy due to the
non-retarded 
BCS coupling $\lambda$. $\hat{\tau}_{i}$ denote Pauli
matrices. $\delta$ is the contour Dirac delta function, and $\star$
denotes the matrix multiplication in temporal and Nambu indices. The
self-energy in Eq.~(\ref{eq:G}) has contributions from the phonon
and disorder scatterings $\hat{\Sigma}_{{\bf k}}=\hat{\Sigma}_{{\bf k}}^{\text{ph}}+\hat{\Sigma}_{{\bf k}}^{\text{dis}}$:
\begin{align}
\hat{\Sigma}_{{\bf k}}^{\text{ph}}\left(t,t'\right) & =-i\frac{1}{V}\sum_{{\bf k'}}g^{2}D_{{\bf k-k'}}\left(t,t'\right)\hat{\tau}_{3}\hat{G}_{{\bf k'}}\left(t,t'\right)\hat{\tau}_{3},\label{eq:sigma}\\
\hat{\Sigma}_{{\bf k}}^{\text{dis}}\left(t,t'\right) & =-i\frac{1}{2\pi\nu_{0}\tau_{\text{el}}}\sum_{{\bf k'}}\hat{\tau}_{3}\hat{G}_{{\bf k'}}\left(t,t'\right)\hat{\tau}_{3},\label{eq:sigma-1}
\end{align}
where the contour-ordered phonon propagator is defined as $D_{{\bf q}}\left(t,t'\right)=-i\left\langle T_{{\cal C}}\phi_{{\bf q}}\left(t\right)\phi_{-{\bf q}}\left(t'\right)\right\rangle $
and $\tau_{\text{el}}^{-1}=2\pi\nu_{0}\overline{U_{{\bf q}}U_{{\bf -q}}}$
is the elastic scattering rate. In the following we ignore the back-action
effects on the phonon propagator and assume it is in equilibrium state
at the initial temperature. We then proceed with the standard quasiclassical
approximation and average the propagator over the Fermi surface $g_{{\bf q}}^{2}D_{{\bf q}}\left(t,t'\right)\rightarrow\left\langle g_{{\bf q}}^{2}D_{{\bf q}}\left(t,t'\right)\right\rangle {}_{\text{FS}}$.
Under these assumptions, the effective phonon propagator is fully
determined by its spectral density (Eliashberg function) which we
denote as $\alpha^{2}F\left(\omega\right)\equiv2\nu_{0}\text{Im}\left\langle g_{{\bf q}}^{2}D_{{\bf q}}^{R}\left(\omega\right)\right\rangle {}_{\text{FS}}$,
where the $D_{{\bf q}}^{R}\left(\omega\right)$ is the retarded part
phonon propagator. Throughout this work we consider an effective Debye
model $\alpha^{2}F\propto\theta\left(\omega_{\text{D}}-\omega\right)\omega^{2}$,
where $\omega_{\text{D}}$ is the cut-off frequency of the phononic band. The
effective electron-phonon pairing strength and the mean frequency
are conventionally defined as $\lambda_{\text{el-ph}}=\int\frac{d\omega}{\omega}\alpha^{2}F\left(\omega\right)$,
$\bar{\omega}=\lambda_{\text{el-ph}}^{-1}\int d\omega\alpha^{2}F\left(\omega\right)$.
The time-dependent order parameter, $\Delta\left(t\right)$, satisfies
the self-consistency equation by replacing the phonon propagator in
Eq.~(\ref{eq:sigma}) with $\lambda\nu_{0}^{-1}\delta\left(t,t'\right)$.
In order to simplify the consideration and avoid the complications
due to the energy-dependent electronic density of states, we assume
the metal is two-dimensional with the high-energy cut-off such
that $\xi_{{\bf k}}\in\left[-E_{F},E_{F}\right]$ \citep{BL06}. We
discuss a caveat related to such simplified dispersion in the SM. 

\paragraph{Electromagnetic response}

We now compute the electromagnetic response of the system. Provided
that the electron-electron interaction is short-range (contact), it
is sufficient to take into account only the mean-field-level contributions.
The gauge-invariant current density to first order in the external
gauge potential ${\bf A}$ comprises paramagnetic and diamagnetic
contributions ${\bf j}_{s}(t)={\bf j}_{\text{p}}(t)+{\bf j}_{\text{d}}(t)$,
where

\begin{align}
{\bf j}_{\text{p}}\left(t\right) & =i\frac{{\bf A}}{V}\sum_{{\bf k}}\frac{{\bf k}^{2}}{dm^{2}}\int_{{\cal C}}ds\text{Tr}\left\{ \hat{G}_{{\bf k}}\left(t,s\right)\hat{G}_{{\bf k}}\left(s,t\right)\right\} ,\label{eq:jp}\\
{\bf j}_{\text{d}}\left(t\right) & =i\frac{{\bf A}}{V}\sum_{{\bf k}}\frac{1}{m}\text{Tr}\left\{ \hat{\tau}_{3}\hat{G}_{{\bf k}}\left(t,t+0^{+}\right)\right\} .\label{eq:jd}
\end{align}
Here the Green's function are taken in the limit ${\bf A}\rightarrow0$
and $d$ is the dimensionality of the problem. We note that the gauge
potential is assumed to be static in Eqs.~(\ref{eq:jp}, \ref{eq:jd})
for simplicity. The time-dependent superfluid density can be found
from Eqs.~(\ref{eq:jp}, \ref{eq:jd}) ${\bf j}\left(t\right)\equiv-\frac{n_{s}\left(t\right)}{m}{\bf A}$.
Before discussing the results of a complete numerical calculation,
we note that in the absence of disorder or phonon scatterings, the
superfluid density does not depend on time to lowest order in $\Delta/E_{F}$.
This is follows from the analysis of the BdG Hamiltonian in
the presence of a static gauge potential $\hat{h}_{{\bf k}}\left({\bf A}\right)\approx(\xi_{{\bf k}}\hat{\tau}_{3}-\frac{{\bf k}{\bf A}}{m}\hat{\tau}_{0}+\frac{{\bf A}^{2}}{2m}\hat{\tau}_{3})+\Delta\left(t\right)\hat{\tau}_{1}$.
Paramagnetic term commutes with the rest of the Hamiltonian and its contribution
to the response given by Eq.~(\ref{eq:jp}) is time-independent.
This implies that any dynamics of the superfluid density requires
a momentum exchange mechanism as schematically shown in Fig~\ref{Fig1}~(a). In our equations of motion Eq.~(\ref{eq:G}),
both self-energies Eqs.~(\ref{eq:sigma}, \ref{eq:sigma-1}) account
for such a mechanism leading to the time-dependent $n_{s}$ as shown
in Fig.~\ref{Fig1}~(b, c). 

We first discuss the phonon-induced scattering case. By varying the
Debye cut-off frequency for a fixed electron-phonon coupling strength
$\lambda_{\text{el-ph}}$, we observe a qualitatively different behavior
of $n_{s}$. In particular, when $\omega_{D}$ is sufficiently large,
both $n_{s}(t)$ and $\Delta\left(t\right)$ are approaching asymptotic
values at large times, consistent with  thermalization
due to coupling to the phonon bath. However, when the non-equilibrium
quasiparticle gap $2\Delta$ is greater than $\omega_{D}$ we see
a different behavior with $n_{s}$ becoming negative. This configuration
is also characterized by the forbidden quasiparticle recombination
which hinders thermalization \cite{CEK23}. Similar phenomena are also observed for
the disordered case for sufficiently large initial and final values
of the BCS pairing strength. We also note that the oscillations of
the superfluid density, which reflect the behavior of $\Delta\left(t\right)$,
are more apparent in the case of a disordered superconductor. In
the remainder of this Letter, we provide a qualitative explanation
to these features. 

\begin{figure}
\includegraphics[scale=0.4]{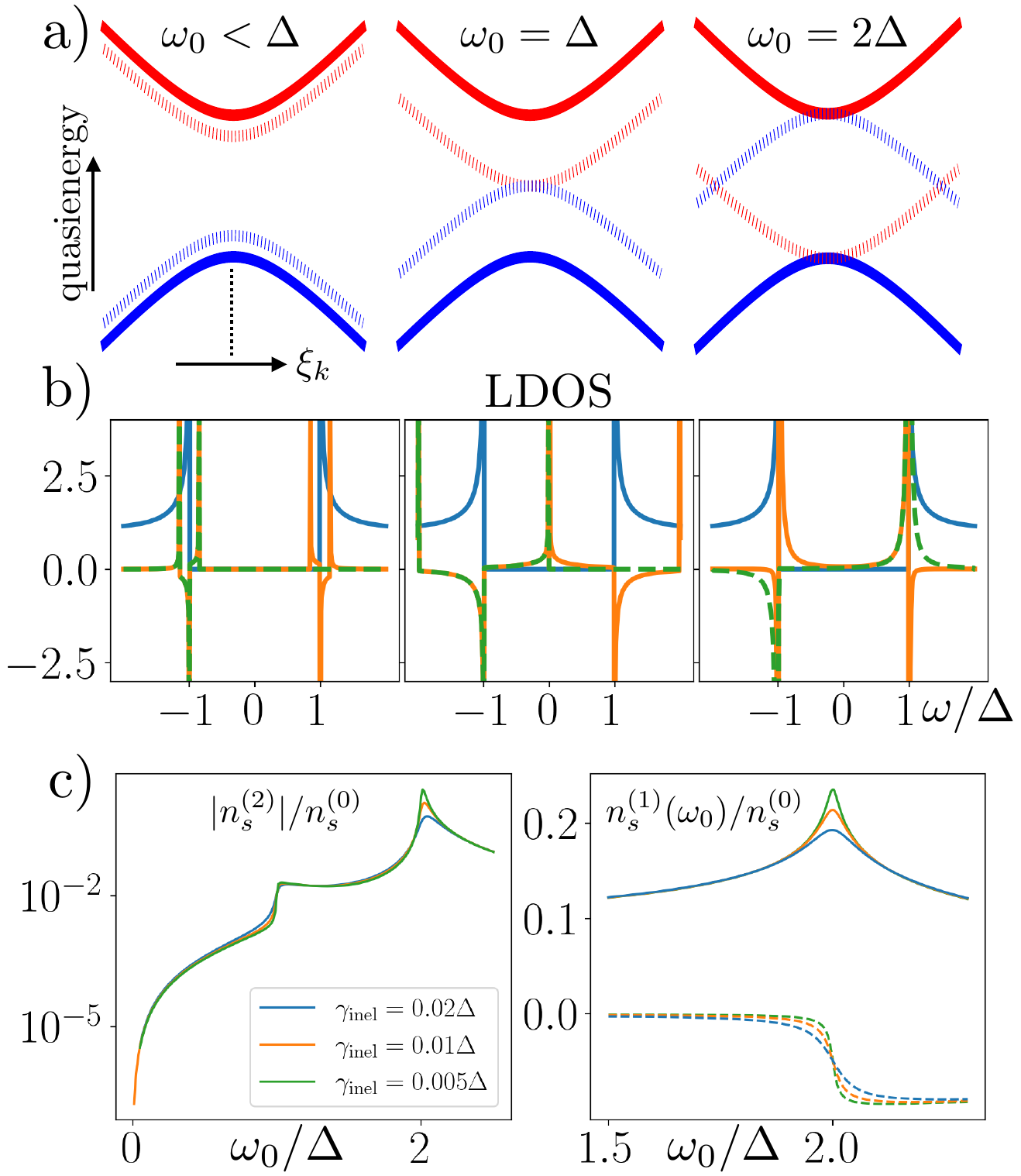}

\caption{Dynamic superfluid density within the Usadel-Keldysh approach. (a)
Schematic representation of Floquet bands, induced by the periodic
gap oscillations at different frequencies. Solid and dashed curves
correspond to the Floquet index $m=0$ and $m=1$ respectively. (b)
Floquet local density of states (LDOS) of physical fermions $\text{Re}\hat{g}_{0,0}^{R}\left(\omega\right)$:
blue and orange curves represent the unperturbed and the second-order
in $\theta$ contributions (shown not to scale). Dashed green curve
represents the LDOS of the occupied Fermionic states defined as $\text{Re}\hat{g}_{0,0}^{<}\left(\omega\right)/2$
\citep{QH17} to second order in $\theta$ (shown not to scale). The
panels from left to right correspond to different gap oscillation
frequencies $\omega_{0}=0.15\Delta$, $\omega_{0}=\Delta$ and $\omega_{0}=2\Delta$.
(c) Superfluid density in diffusive limit for $\theta=0.2\Delta$.
Left and right panels depict static and dynamic components respectively.}

\label{Fig2}
\end{figure}

\begin{figure}
\includegraphics[scale=0.18]{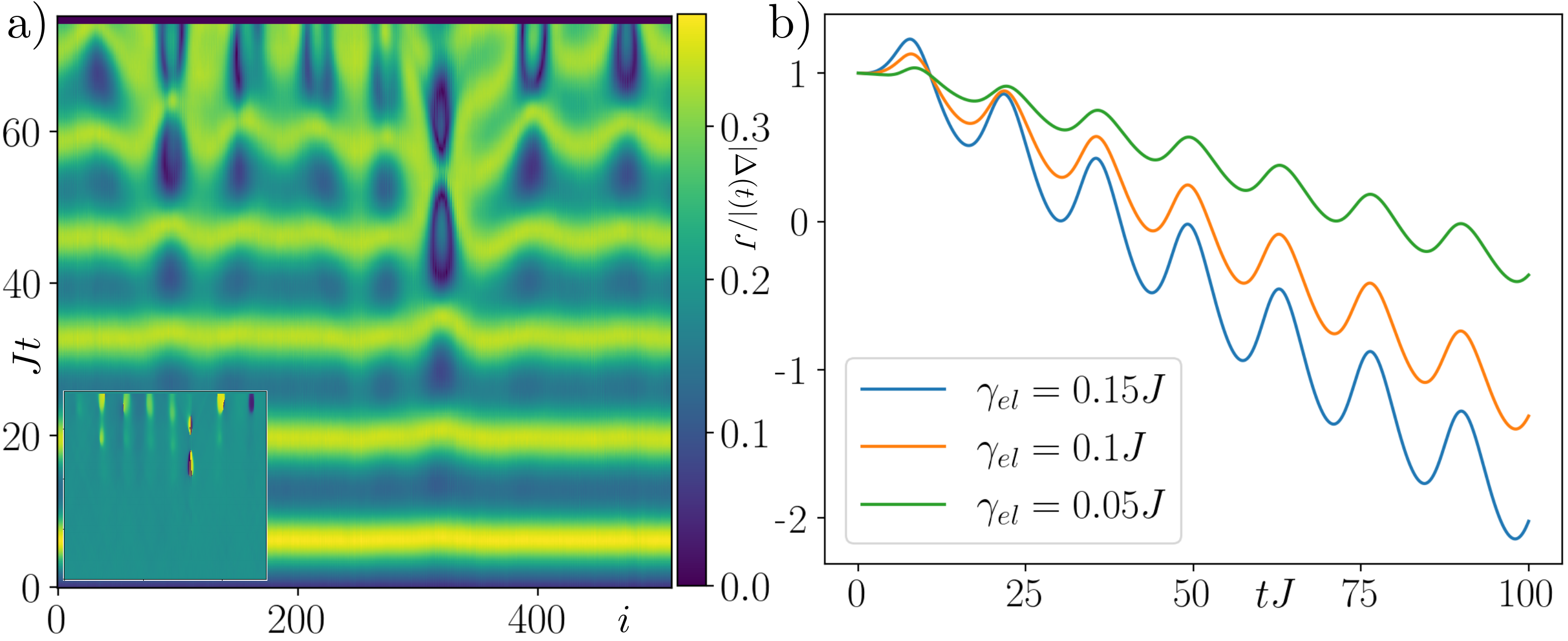}

\caption{Quench dynamics a 1-dimensional disordered superconductor. (a) Time
dependence of the spatially-non-uniform gap in a 1-dimensional Hubbard
model after the quench with $\lambda_{i}=1.2,$ $\lambda_{f}=2$.
Inset shows the evolution of the local phase profile. The effective
elastic scattering rate is $\gamma_{\text{el}}=0.1J$. (b) Numerical
calculation of the superfluid density in a spatially-unresolved (uniform
case) tight-binding model using mean-field equations Eqs.~(\ref{eq:G}-\ref{eq:sigma-1}).}

\label{Fig3}
\end{figure}

\paragraph{Floquet-Usadel description}

Let us now consider a simplified scenario where the superconducting
gap is periodically modulated as $\Delta\left(t\right)=\Delta+\theta\cos\omega_{0}t$,
with $\theta$ and $\omega_{0}$ denoting the amplitude
and the frequency of the gap oscillations respectively. In this case, we can apply the Floquet theory and find the quasienergy spectrum of the time-dependent
BdG Hamiltonian $\hat{h}_{{\bf k}}(t)$, shown schematically in Fig.~\ref{Fig2}~(a)
for different values of $\omega_{0}$. At $\omega_{0}\approx2\Delta$
we observe a crossing of different Floquet bands implying the possibility
of a nearly-resonant excitation of quasiparticles with $\xi_{{\bf k}}\approx0$.
Note that the resonant condition is never satisfied exactly since
for $\xi_{{\bf k}}=0$ the BdG Hamiltonian commutes with its oscillatory
contribution.

In order to find the electromagnetic response of the Higgs-driven
system we apply the quasiclassical approximation to Eq.~(\ref{eq:G})
in the diffusive limit, i.e. assuming that $\tau_{\text{el}}$ is the shortest
timescale of the problem (except for the inverse Fermi energy). Furthermore,
we focus on a steady-state regime of the driving. We consider the Keldysh contour and define $\hat{g}_{t,t'}\left({\bf n}_{k}\right)=\frac{i}{\pi}\int d\xi_{k}\hat{\tau}_{3}\hat{G}_{{\bf k}}\left(t,t'\right)$,
where ${\bf n}_{k}={\bf k}/k$ \citep{LO69}. Following the conventional
approach \citep{LO75}, we represent the quasiclassical Green's function
as $\check{g}_{t,t'}=\left(\begin{array}{cc}
\hat{g}_{t,t'}^{\text{R}} & \hat{g}_{t,t'}^{\text{K}}\\
0 & \hat{g}_{t,t'}^{\text{A}}
\end{array}\right),$ where $\hat{g}_{t,t'}^{\text{R,A,K}}$ are the retarded, advanced
and Keldysh components. In the spatially-uniform case the Green's
function is independent of ${\bf n}_{k}$ and obeys:

\begin{align}
 & \partial_{t}\hat{\tau}_{3}\check{g}_{t,t'}+\partial_{t'}\check{g}_{t,t'}\hat{\tau}_{3}=\left[\Delta\left(t\right)\hat{\tau}_{1}\overset{\star}{,}\check{g}_{t,t'}\right]_{t,t'}+i[\check{\Sigma}^{\text{ph}}\overset{\star}{,}\check{g}]_{t,t'}\label{eq:Usadel}
\end{align}
where $\check{\Sigma}^{\text{ph}}$ is the self-energy due to the
phonon scattering equivalent to Eq.~(\ref{eq:sigma}). The quasiclassical
GF obeys the conventional normalization condition $\check{g}_{t,s}\star\check{g}_{s,t'}=\check{I}\delta\left(t-t'\right),$
where $\check{I}$ is a 4x4 identity matrix. We note that in the uniform
case the self-energy due to disorder scattering commutes with the
Green's function and, therefore, does not contribute to Eq.~(\ref{eq:Usadel}).
To simplify our description, we approximate $\check{\Sigma}^{\text{ph}}$
in the temporal Fourier space as follows \citep{S20,VHB24,DGH84,OHV21}: 

\begin{equation}
\check{\Sigma}_{\omega,\omega'}^{\text{ph}}=\gamma_{\text{inel}}\left(\begin{array}{cc}
\hat{\tau}_{3} & 2\hat{\tau}_{3}\tanh\frac{\beta\omega}{2}\\
0 & -\hat{\tau}_{3}
\end{array}\right)\delta\left(\omega-\omega'\right).
\end{equation}
where $\gamma_{\text{inel}}$ is an effective inelastic scattering
rate and $\beta$ is the inverse temperature. We note that this form
of $\check{\Sigma}^{\text{ph}}$ effectively describes coupling
to a non-superconducting fermionic reservoir. As a result, $\check{\Sigma}^{\text{ph}}$
explicitly breaks the fermion number conservation. However, within
the achievable numerical precision, the density of states and the
average number of fermions are exactly conserved in our model due
to the particle-hole symmetry. In the following, we consider the limit
when the dissipation is weak $\gamma_{\text{inel}}\rightarrow0$. 

The external gauge field can be added to Eq.~(\ref{eq:Usadel}) perturbatively,
yielding the current density ${\bf j}_{t}=i\frac{\sigma_{{\bf N}}}{8}{\bf A}\text{Tr}\left\{ \hat{\tau}_{3}\check{g}\star\left[\hat{\tau}_{3},\check{g}\right]\right\} _{t,t}^{\text{K}}$ \citep{BWB99}, 
where $\sigma_{{\bf N}}=4\pi D\nu_{0}$ is the normal state conductivity,
$D=v_{F}^{2}\tau_{\text{el}}/3$ is the diffusion constant, $\nu_{0}$
is the fermionic density of states, $v_{F}$ is the Fermi velocity
and the superscript $\text{K}$ denotes taking the Keldysh component.
In equilibrium we straightforwardly find $n_{s}=\frac{m}{2}\sigma_{\text{N}}\Delta\tanh\frac{\beta\Delta}{2}$.
Our goal is now to find the corrections to the superfluid density
due to the time dependence of the gap. This is readily achieved by
performing the Fourier transformation of both sides in Eq.~(\ref{eq:Usadel})
with respect to $t$ and $t'$. The resulting equation can be expanded
in powers of $\theta$ and solved iteratively up to the second order, 
e.g. in Mathematica. In Fig.~\ref{Fig3}~(b) we provide the time-averaged
Fermionic local densities of states. As expected, they reflect the
Floquet spectrum of the time-dependent BdG Hamiltonian, schematically
shown in Fig.~\ref{Fig3}~(a). The first order in $\theta$ solution
encodes the oscillating component of the superfluid density $n_{s}^{\left(1\right)}(\omega)$,
shown in Fig.~\ref{Fig3}~(c). For $\omega_{0}\approx2\Delta$ we
observe a logarithmic singularity $n_{s}^{\left(1\right)}/n_{s}^{\left(0\right)}\propto\theta/\Delta\log(\gamma_{\text{inel}}/\Delta)$.
The lowest-order static correction to $n_{s}$ requires expansion
up to the second order in $\theta$ (see also SM for detailed analysis)
which can be performed numerically. For $\omega_{0}\approx2\Delta$
we find the correction having the following scaling $n_{s}^{\left(2\right)}/n_{s}^{\left(0\right)}\propto-\theta^{2}/(\Delta\gamma_{\text{inel}})$
in the limit $\gamma_{\text{inel}}\rightarrow0$ for $\omega_{0}\approx2\Delta$. 

Let us now also discuss the heating of the superconductor due to the
steady-state driving of Higgs oscillations. According to a simple
estimate, provided in the Supplementary material, the melting of the
order parameter can be estimated as the second-order correction to
the static gap $\Delta$, that scales as $\Delta^{\left(2\right)}\propto\theta^{2}/\sqrt{\Delta\gamma_{\text{inel}}}$
with $\Delta^{\left(2\right)}\ll\Delta$. We note that the same scaling
can be expected for the number of excited quasiparticles in the steady
state. Defining the corresponding energy scale as $E_{\text{ex}}=\theta^{2}/\sqrt{\Delta\gamma_{\text{inel}}}$,
we find $n_{s}^{(2)}/n_{s}^{\left(0\right)}\approx-E_{ex}/\sqrt{\Delta\gamma_{\text{inel}}}\rightarrow-\infty$
and $n_{s}^{\left(1\right)}\left(2\Delta\right)\rightarrow0$ in the
limit $\gamma_{\text{inel}}\rightarrow0$. This implies the negative
superfluid density is not induced by the heating but is rather due
to the divergent DOS of Bogolyubov quasiparticles. We also note that for finite but small detuning $\delta=2\Delta-\omega_0>0$ we can take the limit $\gamma_{\text{inel}}\rightarrow0$ explicitly. In this case, the scalings of $n^{(2)}_s$ and $\Delta^{(2)}$ are found to be the same with the replacement $\gamma_{\text{inel}}\rightarrow\delta$.

\paragraph{Quasi-one-dimensional case}

We now discuss the physical implications of the negative superfluid
density. To this end, we consider a superconductor with a spatially-inhomogeneous
order parameter. In order to make the problem tractable, we consider
a disordered one-dimensional tight-binding model with  random on-site
disorder (see SM for more details). In this case the bare electronic
dispersion is given by the conventional expression $\xi_{k}=-2J\cos k$
with $k\in\left[-\pi,\pi\right]$, where $J$ is the nearest-neighbor
tunneling rate. We then numerically self-consistently solve the dynamical
BdG equations motion for $N=512$ sites. The result of the simulation
is shown in Fig.~\ref{Fig3}~(a). We observe the formation of topological
defects in the form of phase slips. In Fig.~\ref{Fig3}~(b) we provide
the result of numerical simulation of the superfluid density in a
uniform one-dimensional tight-binding model using Eqs (\ref{eq:G},
\ref{eq:sigma-1}). We find a good agreement between the
times when the superfluid density becomes negative and when the defects
start to proliferate.

\paragraph{Conclusions and outlook}

In this work we demonstrated that the superfluid density in an amplitude-driven
superconductor can become negative, resulting in proliferation of
spatial inhomogeneities of the order parameter. Our analysis is applicable
to both  superconductors and ultracold Fermi gases. In the
latter case, the Higgs oscillations can be induced by time-dependent
magnetic field with the frequency matching $2\Delta$. We note that in static magnetic field
as a consequence of the fluctuations in $n_{s}$, the superconductor
is expected to emit photons at the frequency $2\Delta$ \cite{MVE17}. The analysis in this Letter is limited to the $s$-wave symmetry of the order parameter and the analysis of more exotic configurations is left for future investigations.   
\paragraph{Acknowledgements.}  This work was supported by the National Science Foundation under Grant No. DMR-2037158, Army Research Office under Grant Number W911NF-23-1-0241, and the Julian Schwinger Foundation. We thank A. Cavalleri, A. Millis, S. Chattopadhyay,  D. Golez, C. Laumann and B. Spivak for useful discussions.

\bibliographystyle{apsrev4-1}
\bibliography{biblio}

\appendix

\section{Floquet-Usadel equations}

In this section we provide details on the solution of the Usadel equation.
By transforming Eq.~(\ref{eq:Usadel}) to Fourier space we get 

\begin{align}
 & -i\omega\hat{\tau}_{3}\check{g}_{\omega,\omega'}+i\omega'\check{g}_{\omega,\omega'}\hat{\tau}_{3}\nonumber \\
= & \int d\Omega\left\{ \Delta\left(\omega-\Omega\right)\hat{\tau}_{1}\check{g}_{\Omega,\omega'}-\check{g}_{\omega,\Omega}\Delta\left(\Omega-\omega'\right)\hat{\tau}_{1}\right\} \nonumber \\
 & +i\check{\Sigma}_{\omega}^{\text{ph}}\check{g}_{\omega,\omega'}-i\check{g}_{\omega,\omega'}\check{\Sigma}_{\omega'}^{\text{ph}}\label{eq:Floquet}
\end{align}
where $\Delta\left(\Omega\right)\equiv\frac{1}{2\pi}\int dte^{i\Omega t}\Delta\left(t\right)$.
For $\Delta\left(t\right)=\Delta+\theta\cos\omega t$ we get $\Delta\left(\Omega\right)=\delta\left(\Omega\right)+\frac{\theta}{2}(\delta(\Omega+\omega)+\delta(\Omega-\omega))$.
Eq.~(\ref{eq:Floquet}) needs to be solved iteratively at least up
to the second order in $\theta$ as it is the lowest non-vanishing
order, contributing to the DC superfluid density. 

\subsection{Supercurrent }

\begin{figure}
\includegraphics[scale=0.38]{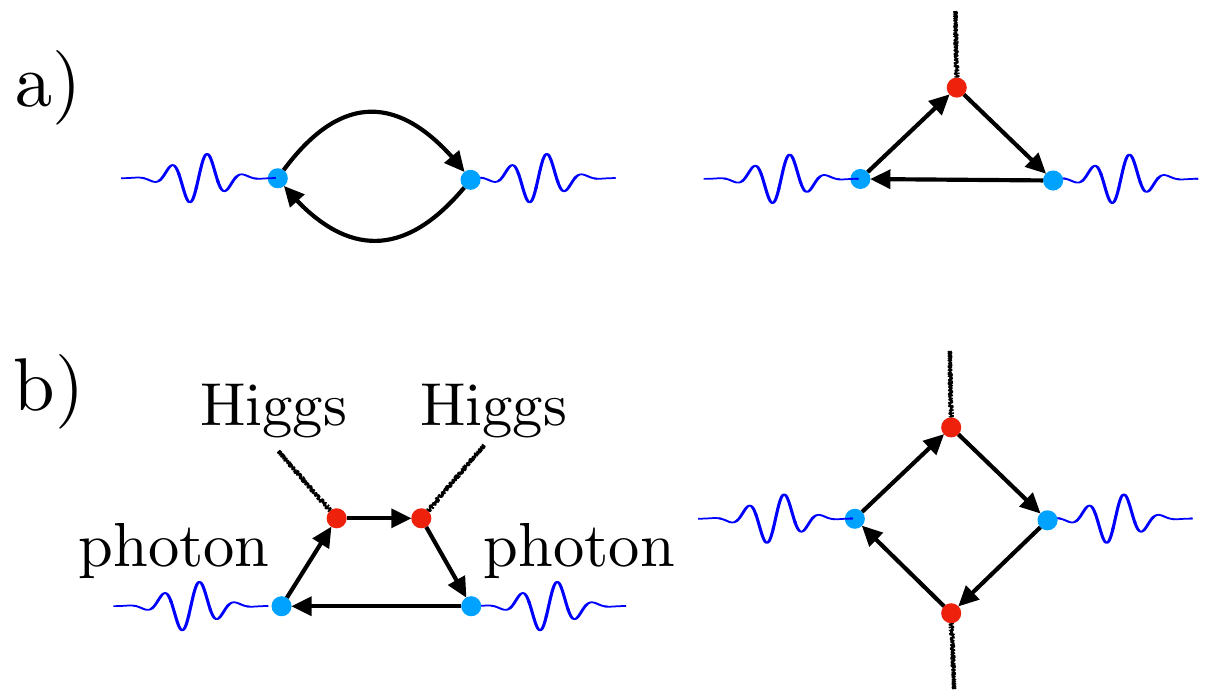}

\caption{Diagrams contributing to the dynamical superfluid density up to the
second order in $\theta$. (a) Zeroth order (equilibrium) Eq.~\eqref{eq:Jo} and first
order Eq.~\eqref{eq:j1}, describing the oscillating part of $n_{s}$. (b) Contribution
to the static component of the superluid density in second order in
$\theta$ Eq.~\eqref{eq:j2}. Blue and red circles respectively represent the current
($\hat{\tau}_{0}$) and the Higgs ($\hat{\tau}_{1}$) vertices respectively.}

\label{Fig4}
\end{figure}

Once the Green's function is determined, we can find its effect on
the supercurrent \citep{BWB99}. 

\begin{align}
{\bf j}_{t} & =i\frac{\sigma_{{\bf N}}}{8}{\bf A}\int ds\text{Tr}\left\{ \hat{\tau}_{3}\hat{g}_{t,s}\hat{\tau}_{3}\hat{g}_{s,t}\right\} ^{\text{K}},\label{eq:jt}
\end{align}
where we used the normalization condition of the quasiclassical GF.
We can explicitly write the Keldysh component of the product in Eq.~(\ref{eq:jt}):

\begin{align*}
{\bf j}_{t} & =i\frac{\sigma_{{\bf N}}}{8}{\bf A}\int ds\text{Tr}\left\{ \hat{\tau}_{3}\hat{g}_{t,s}^{\text{R}}\hat{\tau}_{3}\hat{g}_{s,t}^{\text{K}}\right\} \\
 & +i\frac{\sigma_{{\bf N}}}{8}{\bf A}\int ds\text{Tr}\left\{ \hat{\tau}_{3}\hat{g}_{t,s}^{\text{K}}\hat{\tau}_{3}\hat{g}_{s,t}^{\text{A}}\right\} .
\end{align*}
Let us now take Fourier transform of both sides:

\begin{align}
{\bf j}_{\Omega}\equiv & \int\frac{dt}{2\pi}e^{i\Omega t}{\bf j}_{t}=\nonumber \\
 & i\frac{\sigma_{{\bf N}}}{8}{\bf A}\int\frac{d\omega}{2\pi}\int d\omega'\text{Tr}\left\{ \hat{\tau}_{3}\hat{g}_{\omega',\omega}^{\text{R}}\hat{\tau}_{3}\hat{g}_{\omega,\omega'-\Omega}^{\text{K}}\right\} \\
 & +i\frac{\sigma_{{\bf N}}}{8}{\bf A}\int\frac{d\omega}{2\pi}\int d\omega'\text{Tr}\left\{ \hat{\tau}_{3}\hat{g}_{\omega',\omega}^{\text{K}}\hat{\tau}_{3}\hat{g}_{\omega,\omega'-\Omega}^{\text{A}}\right\} \label{eq:Jo}
\end{align}
By expanding to first order in $\theta$ we get:

\begin{align}
{\bf j}_{\Omega}^{\left(1\right)} & =i\frac{\sigma_{{\bf N}}}{8}{\bf A}\int\frac{d\omega}{2\pi}\text{Tr}\left\{ \hat{\tau}_{3}\hat{g}_{\Omega+\omega,\omega}^{\left(1\right)\text{R}}\hat{\tau}_{3}\hat{g}_{\omega}^{\left(0\right)\text{K}}\right\} \nonumber \\
 & +i\frac{\sigma_{{\bf N}}}{8}{\bf A}\int\frac{d\omega}{2\pi}\text{Tr}\left\{ \hat{\tau}_{3}\hat{g}_{\omega}^{\left(0\right)\text{R}}\hat{\tau}_{3}\hat{g}_{\omega,\omega-\Omega}^{\left(1\right)\text{K}}\right\} \nonumber \\
 & +i\frac{\sigma_{{\bf N}}}{8}{\bf A}\int\frac{d\omega}{2\pi}\text{Tr}\left\{ \hat{\tau}_{3}\hat{g}_{\Omega+\omega,\omega}^{\left(1\right)\text{K}}\hat{\tau}_{3}\hat{g}_{\omega}^{\left(0\right)\text{A}}\right\} \nonumber \\
 & +i\frac{\sigma_{{\bf N}}}{8}{\bf A}\int\frac{d\omega}{2\pi}\text{Tr}\left\{ \hat{\tau}_{3}\hat{g}_{\omega}^{\left(0\right)\text{K}}\hat{\tau}_{3}\hat{g}_{\omega,\omega-\Omega}^{\left(1\right)\text{A}}\right\} ,\label{eq:j1}
\end{align}
where we used that the zeroth-order Green's functions are translationally-invariant
in time. The superscript $\ldots^{\left(i\right)}$ denotes the order
in expansion in $\theta$. Using Eq.~(\ref{eq:Floquet}) we explicitly
see that the lowest order term is oscillatory. Let us now consider
the second-order contribution. Moreover, we will be interested in
a DC component which is found by setting $\Omega=0$ in Eq.~(\ref{eq:Jo}).
We get: 

\begin{align}
{\bf j}_{\Omega=0}^{\left(2\right)} & =i\frac{\sigma_{{\bf N}}}{8}{\bf A}\int d\omega\int d\omega'\text{Tr}\left\{ \hat{\tau}_{3}\left(\hat{g}_{\omega,\omega'}^{\left(1\right)\text{R}}+\hat{g}_{\omega,\omega'}^{\left(1\right)\text{A}}\right)\hat{\tau}_{3}\hat{g}_{\omega',\omega}^{\left(1\right)\text{K}}\right\} \nonumber \\
 & +i\frac{\sigma_{{\bf N}}}{8}{\bf A}\int d\omega\int d\omega'\text{Tr}\left\{ \hat{\tau}_{3}\left(\hat{g}_{\omega,\omega'}^{\left(0\right)\text{R}}+\hat{g}_{\omega,\omega'}^{\left(0\right)\text{A}}\right)\hat{\tau}_{3}\hat{g}_{\omega',\omega}^{\left(2\right)\text{K}}\right\} \nonumber \\
 & +i\frac{\sigma_{{\bf N}}}{8}{\bf A}\int d\omega\int d\omega'\text{Tr}\left\{ \hat{\tau}_{3}\left(\hat{g}_{\omega,\omega'}^{\left(2\right)\text{R}}+\hat{g}_{\omega,\omega'}^{\left(2\right)\text{A}}\right)\hat{\tau}_{3}\hat{g}_{\omega',\omega}^{\left(0\right)\text{K}}\right\} ,\label{eq:j2}
\end{align}
 The two terms are found to be much more singular than the first one.
The resulting terms can be represented diagrammatically as shown in
Fig.~\ref{Fig4}. The last two two terms can be simplified even further
by using the fact that the unperturbed Green's functions are time-translationally
invariant, i.e. $\propto\delta\left(\omega-\omega'\right)$.

\section{Para- and diamagnetic terms}

In this section, discuss a caveat that arises from our simplified band structure.
Consider a linear response in equilibrium in a normal state of a generic
interacting electronic band with dispersion $\xi_{{\bf k}}$. The paramagnetic susceptibility reads: 

\begin{equation}
\Pi_{i,i}\left(i\Omega_{m}\right)=\frac{1}{\beta V}\sum_{{\bf k}}\left(\frac{\partial\xi_{{\bf k}}}{\partial k_{i}}\right)^{2}{\cal G}_{{\bf k}}\left(i\epsilon_{n}+i\Omega_{m}\right){\cal G}_{{\bf k}}\left(i\epsilon_{n}\right)\label{eq:Pii}
\end{equation}
where ${\cal G}_{{\bf k}}\left(i\epsilon_{n}\right)$ is the Matsubara
Green's function with $\epsilon_{n}=(2n+1)\pi/\beta$, $n\in\mathbb{Z}$.
Note that Eq.~(\ref{eq:Pii}) is valid only when the interaction
and the disorder are short-range. Performing the analytic continuation
we get:

\[
\Pi_{i,i}^{R}\left(0\right)=\Im\int\frac{dx}{2\pi}\frac{d{\bf k}}{\left(2\pi\right)^{2}}\left(\frac{\partial\xi_{{\bf k}}}{\partial k_{i}}\right)^{2}\left(G_{{\bf k}}^{R}\left(x\right)\right)^{2}\tanh\left(\frac{\beta x}{2}\right),
\]
where $G_{k}^{R}\left(x\right)=\left(x-\xi_{k}-\Sigma^{R}\left(x\right)\right)^{-1}$
is retarded Green's function and $\Sigma^{R}$ is the self-energy.
We note that the independence of $\Sigma^{R}$ of momentum is consistent
with the usage of unrenormalized current vertices in Eq.~(\ref{eq:Pii}).
Let us transform this expression as follows:

\begin{align*}
\Pi_{i,i}^{R}\left(0\right) & =\Im\int\frac{dx}{2\pi}\frac{d{\bf k}}{\left(2\pi\right)^{2}}\left(\frac{\partial\xi_{{\bf k}}}{\partial k_{i}}\right)^{2}\partial_{\xi_{{\bf k}}}G_{{\bf k}}^{R}\left(x\right)\tanh\left(\frac{\beta x}{2}\right)\\
 & =\Im\int\frac{dx}{2\pi}\frac{d{\bf k}}{\left(2\pi\right)^{2}}\frac{\partial\xi_{{\bf k}}}{\partial k_{i}}\partial_{k_{i}}G_{{\bf k}}^{R}\left(x\right)\tanh\left(\frac{\beta x}{2}\right).
\end{align*}
Integrating by parts we get:

\begin{align}
\Pi_{i,i}^{R}\left(0\right) & =\Im\int\frac{dx}{2\pi}\frac{d{\bf k}}{\left(2\pi\right)^{2}}\partial_{k_{i}}\left\{ \frac{\partial\xi_{{\bf k}}}{\partial k_{i}}G_{{\bf k}}^{R}\left(x\right)\tanh\left(\frac{\beta x}{2}\right)\right\} \nonumber \\
 & -\Im\int\frac{dx}{2\pi}\frac{d{\bf k}}{\left(2\pi\right)^{2}}\frac{\partial^{2}\xi_{{\bf k}}}{\partial k_{i}^{2}}G_{{\bf k}}^{R}\left(x\right)\tanh\left(\frac{\beta x}{2}\right).
\end{align}
The first term should vanish for all physical bands while the second
is equal to the diamagnetic contribution. Indeed let us now use the
fact $\text{\ensuremath{\tanh\frac{\beta x}{2}=1-2n_{F}\left(x\right)}}$,
where $n_{F}$ is Fermi-Dirac distribution function and $\int dxG_{k}^{R}\left(x\right)=-\pi$
as a consequence of fermionic the anti-commutation relations:

\begin{align}
\Pi_{i,i}^{R}\left(0\right) & =\frac{1}{2}\int\frac{d{\bf k}}{\left(2\pi\right)^{2}}\frac{\partial^{2}\xi_{{\bf k}}}{\partial k_{i}^{2}}\nonumber \\
 & +\Im\int\frac{dx}{\pi}\frac{d{\bf k}}{\left(2\pi\right)^{2}}\frac{\partial^{2}\xi_{{\bf k}}}{\partial k_{i}^{2}}G_{{\bf k}}^{R}\left(x\right)n_{F}\left(x\right)
\end{align}
Again, the first term is zero since $\partial\xi_{{\bf k}}/\partial k_{i}$
is vanishing at the Brillouin zone edge. Let us repeat the same calculation
for our simplified band $\xi_{k}=k^{2}/2m-\mu$ with a high-energy
cut-off such that $\xi_{k}\in\left[-E_{F},E_{F}\right]$. We observe
that for $\xi_{k}\sim E_{F}$ the velocity vertices do not vanish.
Let us summarize all unphysical contribution to the polarization:

\begin{align*}
\tilde{\Pi}_{i,i}^{R} & =-\int\frac{d{\bf k}}{\left(2\pi\right)^{2}}\partial_{k_{i}}\left\{ \frac{\partial\xi_{{\bf k}}}{\partial k_{i}}n_{{\bf k}}\right\} \\
 & =-\frac{1}{2}\int\frac{d{\bf k}}{\left(2\pi\right)^{2}}\nabla_{{\bf k}}{\bf F}\left({\bf k}\right),
\end{align*}
where ${\bf F}\left({\bf k}\right)=\nabla\cdot\xi_{{\bf k}}n_{{\bf k}}$.
Using Ostrogradsky-Gauss theorem we can find

\begin{align*}
\tilde{\Pi}_{i,i}^{R} & =-\frac{1}{\left(2\pi\right)}E_{F}n_{{\bf \tilde{k}}}
\end{align*}
where $\tilde{{\bf k}}$ is the cut-off momentum. Note that this contribution
is exponentially suppressed and is typically negligible in our parameter
range. We however can eliminate this contribution exactly by renormalizing
the velocity vertices, i.e. assuming $\partial_{k_{i}}\xi=\frac{k_{i}}{m}\theta\left(k\right)$,
where $\theta\left(k\right)$ is some high-energy cut-off function
obeying $\theta(\tilde{k})=0$. This is equivalent to a flattening
of the electronic dispersion at high energies. We numerically find
that the exact functional form of $\theta$ is not important and we
can choose it to be a Heaviside theta. 

\section{Quasi-1d superconductor}

Consider the dynamics of an inhomogeneous 1d superconductor. The BdG
Hamiltonian, of a tight-binging superconductor reads:

\begin{align*}
\hat{H}_{\text{BdG}}\left(t\right) & =J\sum_{i}\left(\left|i\right\rangle \left\langle i+1\right|+\left|i+1\right\rangle \left\langle i\right|\right)\otimes\hat{\tau}_{3}\\
 & +\sum_{i}u_{i}\left|i\right\rangle \left\langle i\right|\otimes\hat{\tau}_{3}+\hat{\Sigma}\left(t\right)
\end{align*}
where $u_{i}$ is the local disorder potential, $J$ is the nearest-neighbor
tunneling rate and $|i\rangle$ is the state locating an electron in the cite $i$. The self-energy can be parametrized as $\hat{\Sigma}\left(t\right)=\sum_{i}\left(\Delta_{i}\hat{\tau}_{+}+\Delta_{i}^{*}\hat{\tau}_{-}+\delta u_{i}\hat{\tau}_{3}\right)\left|i\right\rangle \left\langle i\right|$,
$\Delta_{i}$ is the local gap and $\delta u_{i}$ is the renormalized
disorder. Self-consistency equation reads (note that it also has normal
components which slightly renormalize the disorder):

\begin{equation}
\hat{\Sigma}_{i}\left(t\right)=i\frac{\lambda}{2}\hat{\tau}_{3}\hat{G}_{i,i}^{\text{K}}\left(t,t\right)\hat{\tau}_{3},\label{eq:sigm}
\end{equation}
where $\hat{G}^{K}$ is Keldysh Green's function obeying:

\begin{align*}
 & i\partial_{t}\hat{G}^{\text{K}}\left(t,t'\right)-\hat{H}_{\text{BdG}}\left(t\right)\hat{G}^{\text{K}}\left(t,t'\right)=0\\
 & -i\partial_{t'}\hat{G}^{\text{K}}\left(t,t'\right)-\hat{G}^{\text{K}}\left(t,t'\right)\hat{H}_{\text{BdG}}\left(t'\right)=0.
\end{align*}
Equivalently, we can write

\begin{align*}
 & \partial_{t}\hat{G}^{\text{K}}\left(t,t\right)+\partial_{t'}\hat{G}^{\text{K}}\left(t,t'\right)=\\
 & -i\left\{ \hat{H}_{\text{BdG}}\left(t\right)\hat{G}^{\text{K}}\left(t,t'\right)-\hat{G}^{\text{K}}\left(t,t'\right)\hat{H}_{\text{BdG}}\left(t'\right)\right\} .
\end{align*}
By setting $t=t'$ we get:

\begin{equation}
\partial_{t}\hat{G}^{\text{K}}\left(t,t\right)=-i\left[\hat{H}_{\text{BdG}}\left(t\right),\hat{G}^{\text{K}}\left(t,t\right)\right]\label{eq:dyn}
\end{equation}
This equation can be solved numerically self-consistently with Eq.~(\ref{eq:sigm}).
At $t=0$ in equilibrium we have:

\begin{align*}
\hat{G}^{R}\left(\omega\right)= & \sum_{l}\frac{1}{\omega+i0^{+}-\gamma_{l}}\left|l\right\rangle \left\langle l\right|,\\
\hat{G}^{K}\left(\omega\right) & =\left(\hat{G}^{R}\left(\omega\right)-\hat{G}^{A}\left(\omega\right)\right)\tanh\frac{\beta\omega}{2}\\
 & =-2\pi i\sum_{l}\left|l\right\rangle \left\langle l\right|\delta\left(\omega-\gamma_{l}\right)\tanh\frac{\beta\omega}{2},
\end{align*}

where we denoted the eigenvalues and eigenvectors of BdG Hamiltonian
at $t=0$ as $\gamma_{l}$ and $\left|l\right\rangle $. Thus, same
time:

\begin{equation}
\hat{G}^{K}\left(t=0,t=0\right)=i\sum_{l}\left|l\right\rangle \left\langle l\right|\tanh\frac{\beta\gamma_{l}}{2}\label{eq:GK-3}
\end{equation}

Eqs. (\ref{eq:GK-3},\ref{eq:dyn},\ref{eq:sigm}) can be solved self-consistently
numerically. The result of numerical simulation is shown in Fig.~\ref{Fig3}~(a).

\section{Estimation of heating}

The heating can be determined via consistency of our quasiclassical
steady-state description in the main text. Let us now assume the oscillations
of the amplitude mode are induced by the time-dependent BCS interaction
strength $\lambda\left(t\right)=\lambda_{0}\left(1+\lambda_{1}\left(t\right)\right)$,
where $\lambda_{0}$ and $\lambda_{1}$ are static and oscillatory
contributions respectively. Moreover, we assume $\lambda_{1}\ll1$.
The quasiclassical BCS self-consistency equation writes:

\[
\Delta\left(t\right)=\frac{i}{8}\lambda_{0}\left(1+\lambda_{1}\left(t\right)\right)\text{Tr}\hat{\tau}_{1}\bar{g}_{t,t}
\]
where $\bar{g}_{t,t}\equiv\int_{-\omega_{\text{co}}}^{\omega_{\text{co}}}\frac{d\omega}{2\pi}\hat{g}_{t,\omega}^{\text{K}}$
denoting the regularized Keldysh Green's function with the center-of-mass
time $t$ and $\omega$ being the relative frequency. $\omega_{\rm{co}}$ denotes the high-energy cut-off. Expanding in
small $\lambda_{1}$ we get:

\begin{align*}
\Delta^{\left(0\right)}\left(t\right) & =\frac{i}{8}\lambda_{0}\text{Tr}\hat{\tau}_{1}\bar{g}_{t,t}^{\left(0\right)}\\
\Delta^{\left(1\right)}\left(t\right) & =\frac{i}{8}\lambda_{0}\text{Tr}\hat{\tau}_{1}\bar{g}_{t,t}^{\left(1\right)}+\frac{i}{8}\lambda_{0}\lambda_{1}\left(t\right)\text{Tr}\hat{\tau}_{1}\bar{g}_{t,t}^{\left(0\right)}\\
\Delta^{\left(2\right)}\left(t\right) & =\frac{i}{8}\lambda_{0}\text{Tr}\hat{\tau}_{1}\bar{g}_{t,t}^{\left(2\right)}+\frac{i}{8}\lambda_{0}\lambda_{1}\left(t\right)\text{Tr}\hat{\tau}_{1}\bar{g}_{t,t}^{\left(1\right)}
\end{align*}

In our notations $\Delta^{\left(1\right)}\left(t\right)=\theta$$\cos\omega_{0}t$,
$\Delta^{\left(0\right)}=\Delta$. We note that we can always find
$\lambda_{1}\left(t\right)$ that satisfies the self-consistency equation
above for $\Delta^{\left(1\right)}$. This is achieved by choosing

\[
\lambda_{1}\left(t\right)=\frac{\Delta^{\left(1\right)}\left(t\right)}{\Delta}-\frac{i}{8}\frac{\lambda_{0}}{\Delta}\text{Tr}\hat{\tau}_{1}\bar{g}_{t,t}^{\left(1\right)}
\]
 Let us now consider the second-order correction to the gap. 

\begin{align*}
 & \Delta^{\left(2\right)}\left(t\right)=\frac{i}{8}\lambda_{0}\text{Tr}\hat{\tau}_{1}\bar{g}_{t,t}^{\left(2\right)}+\lambda_{1}\left(t\right)\Delta^{\left(1\right)}\left(t\right)-\lambda_{1}^{2}\left(t\right)\Delta
\end{align*}
For consistency, we need $\Delta^{\left(2\right)}\ll\Delta$. By evaluating
$\text{Tr}\hat{\tau}_{1}\bar{g}_{t,t}^{\left(2\right)}$ numerically
and $\Delta^{\left(1\right)}$ using Higgs susceptibility, we find
that the time-averaged $\Delta^{\left(2\right)}$ scales as $\propto\theta^{2}/\sqrt{\Delta\gamma_{\text{inel}}}$
for $\omega_{0}\approx2\Delta$, consistently with our expression for the
heating in the main text. 

Let us provide an alternative estimation (explaining the $1/\sqrt{\gamma_{\text{inel}}}$ scaling) of the heating as the number
of excited quasiparticles due to the driving of the Higgs mode. Let
us start with the case with no disorder. To this end, by performing
a unitary transformation that diagonalizes the static part of the
BdG Hamiltonian we get $U\hat{h}_{{\bf k}}\left(t\right)U^{\dagger}=H_{0}+H_{1}$:

\begin{align*}
 & H_{0}=\sum_{{\bf k}}\lambda_{{\bf k}}\left(c_{{\bf k}}^{\dagger}c_{{\bf k}}+d_{-{\bf k}}^{\dagger}d_{-{\bf k}}\right)\\
 & H_{1}=\sum_{{\bf k}}\theta_{{\bf k}}\left(t\right)c_{{\bf k}}d_{-{\bf k}}+\theta_{{\bf k}}\left(t\right)d_{-{\bf k}}^{\dagger}c_{{\bf k}}^{\dagger},
\end{align*}
where $\left(\begin{array}{c}
c_{{\bf k}}\\
d_{-{\bf k}}^{\dagger}
\end{array}\right)=U\left(\begin{array}{c}
\psi_{{\bf k},\downarrow}\\
\psi_{-{\bf k},\uparrow}^{\dagger}
\end{array}\right)$, $\theta_{{\bf k}}=\theta\cos(\omega_{0}t)\xi_{k}/\sqrt{\xi_{k}^{2}+\Delta^{2}}$,
$\lambda_{k}=\sqrt{\xi_{k}^{2}+\Delta^{2}}$. We note that we ignored
the excitation-conserving terms in $H_{1}$ since they do not affect
the heating. Assuming the system at $t=-\infty$ is prepared in a
non-interacting ground state of the BdG Hamiltonian $\left|\Omega\right\rangle $,
the time evolution is can be represented as:

\begin{align*}
\left|\psi\left(t\right)\right\rangle  & =Te^{-i\int_{-\infty}^{t}H_{1}\left(s\right)ds}\left|\Omega\right\rangle ,
\end{align*}
where $T$ denotes the time-ordering. Expanding up to the first order
in $\theta$ and keeping only the resonant terms we get:

\begin{align*}
 & \left|\psi^{\left(1\right)}\left(t\right)\right\rangle \approx\\
 & -i\theta\sum_{{\bf k}}\frac{\xi_{k}}{\sqrt{\xi_{k}^{2}+\Delta^{2}}}\int_{-\infty}^{t}e^{i\left(\omega_{0}-2\lambda_{k}\right)s}e^{-\left(t-s\right)\gamma_{\text{inel}}}ds\left|1_{{\bf k}}1_{-{\bf k}}\right\rangle \\
 & =-i\theta\sum_{{\bf k}}\frac{\xi_{k}}{\sqrt{\xi_{k}^{2}+\Delta^{2}}}\frac{e^{it\left(\omega_{0}-2\lambda_{k}\right)}}{\gamma_{\text{inel}}-2i\lambda_{k}+i\omega_{0}}\left|1_{{\bf k}}1_{-{\bf k}}\right\rangle 
\end{align*}
where we introduced the pair-breaking parameter $\gamma_{\text{inel}}$
describing thermalization. Let us now compute the number of excited
pairs in this system. Defining $\hat{N}=V^{-1}\sum_{{\bf k}}\left(c_{{\bf k}}^{\dagger}c_{{\bf k}}+d_{-{\bf k}}^{\dagger}d_{-{\bf k}}\right)$
we find

\begin{equation}
\left\langle \hat{N}\right\rangle =2\theta^{2}V^{-1}\sum_{{\bf k}}\frac{\xi_{k}^{2}}{\xi_{k}^{2}+\Delta^{2}}\frac{1}{\gamma_{\text{inel}}^{2}+\left(2\lambda_{k}-\omega_{0}\right)^{2}}.\label{eq:EqNex}
\end{equation}
Upon taking this integral we get $N_{ex}=\left\langle \hat{N}\right\rangle \propto\theta^{2}\nu_{0}/\sqrt{\Delta\gamma_{\text{inel}}}$
for $\omega_{0}=2\Delta$. We note that the same scaling as in Eq.~(\ref{eq:EqNex}) can
also be inferred from linear response by taking the imaginary part
of the Higgs susceptibility within quasiclassical approach. This implies
that our result for $N_{ex}$ does not depend on presence of disorder.

\end{document}